\documentclass[journal, 10pt]{IEEEtran}
\usepackage{epsfig}
\usepackage{subfigure}
\usepackage{amsmath}
\usepackage{indentfirst}
\usepackage{caption}
\usepackage{cite}
\usepackage{setspace}
\usepackage{multicol}

\begin{document}

\title{Cooperative Communication Based on Random Beamforming Strategy in Wireless Sensor Networks}

\author{Li~Li,
        Kamesh~Namuduri,
        Shengli~Fu
        \\ Electrical Engineering Department
        \\ University of North Texas
        \\ Denton, TX 76201
        \\ lili@my.unt.edu, kamesh.namuduri@unt.edu,  fu@unt.edu
        }

\maketitle

%\doublespacing

\begin{abstract}

This paper presents a two-phase cooperative communication strategy and an optimal power allocation strategy to transmit sensor observations to a fusion center in a large-scale sensor network. Outage probability is used to evaluate the performance of the proposed system. Simulation results demonstrate that: 1) when signal-to-noise ratio is low, the performance of the proposed system is better than that of the multiple-input and multiple-output system over uncorrelated slow fading Rayleigh channels; 2) given the transmission rate and the total transmission SNR, there exists an optimal power allocation that minimizes the outage probability; 3) on correlated slow fading Rayleigh channels, channel correlation will degrade the system performance in linear proportion to the correlation level.

\end{abstract}


\IEEEpeerreviewmaketitle

%\doublespacing

\section{Introduction}
\begin{figure}
\begin{center}
\includegraphics[width=3.5in]{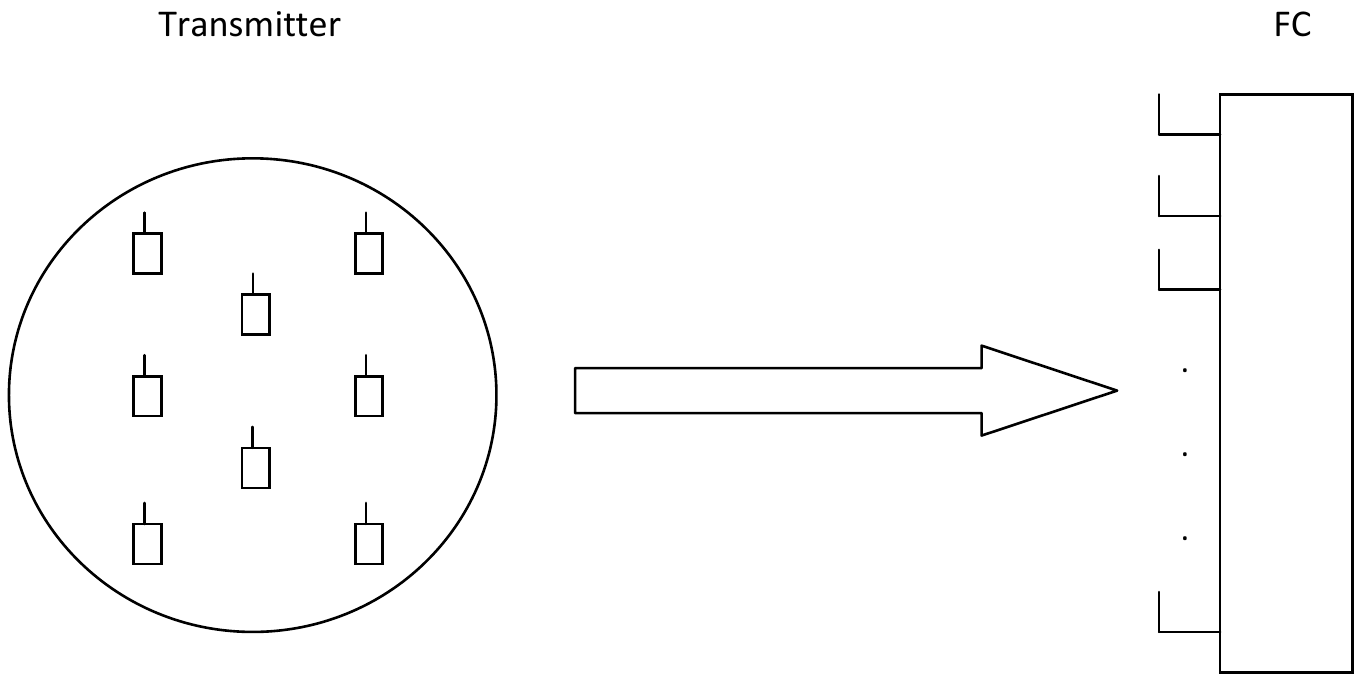}
\caption{Cluster based sensor network model} \label{Fig:mod1}
\end{center}
\end{figure}

\IEEEPARstart{W}{ireless} sensor networks (WSN) received a lot of attention in recent years due to their applications in numerous areas such as environmental monitoring, health care, military surveillance, crisis management, and transportation. They typically consist of a large number of sensing devices organized into a network that is capable of monitoring an environment and reporting the collected data to a fusion center (FC). Fig.\ref{Fig:mod1} shows an example of cluster based sensor network. 

Fading is a typical characteristic of wireless channels, and it is caused by multipath and mobility. One important strategy to combat channel fading is the use of  diversity \cite{Tarokh98} \cite{Tarokh99}, which can be created over time, frequency, and space. The basic idea of obtaining diversity as well as improving the system performance is to create several independent signal paths between the transmitter and the receiver.

There is another form of diversity called multiuser diversity that is focused on the uplink in a single cell \cite{Knopp95}. A multiuser diversity system can improve channel capacity by exploiting fading. In this model, multiple users communicate to the base station on time-varying fading channels, and the receiver will track the channel state information and feed back to the transmitters. An efficiency strategy to maximize the total information-theoretic capacity is to schedule at any one time only the user with the best channel to transmit to the base station. Diversity gain is obtained by finding one among all the independent user channels that is near its peak. It can also be considered as another form of selection diversity.

In wireless sensor networks, the design of optimal sensor deployment strategies is influenced by energy savings. Distributed cooperative communication strategies achieve energy saving through spatial diversity \cite{shuguang04} \cite{Laneman03}. The use of cooperative transmission and/or reception of data among sensors minimizes the per-node energy consumption and thus increases wireless sensor network's lifetime \cite{Laneman04}.

Multiuser diversity is combined with transmit beamforming in \cite{Viswanath02} to achieve \textit{coherent beamforming} capacity. In this model, the transmitter only requires received signal-to-noise ratio (SNR) in the form of feedback. However, this design is based on an assumption of having multiple antennas at the receiving base station. In a cluster based wireless sensor network, however, there are several antennas in each transmitting cluster, which makes opportunistic beamforming method incapable of increasing capacity \cite{Chung03}. In order to cope with this situation, multiplexing is used rather than beamforming in \cite{Chung03}. In this method, the capacity increases linearly as a function of number of transmit antennas. Thus, there is a need to develop new approachs to increase the channel capacity when using random beamforming. 

In this paper, we propose a cluster based cooperative transmission strategy to achieve multiuser diversity using \textit{random beamforming}. We consider a wireless sensor network with a clustered topology with each cluster consisting of several number of sensors. Each sensor in the transmitting cluster is capable of processing the collecting data and transmitting it through its embedded antenna. The receiving cluster is modeled as a single unit with multiple receiving antennas, and is referred to as virtual fusion center (VFC).

The proposed data transmission involves two phases: (1) intra-cluster phase in which sensors within a cluster communicate with each other over a broadcast channel (each node using one time unit to broadcast), and (2) cluster to VFC phase in which all sensors in the transmitting cluster communicate with the VFC using beamforming. If we consider VFC as a receiving cluster, then, the idea is similar to transmitting data from one cluster to another \cite{Laneman04}. The VFC combines the received data using maximal ratio combining (MRC) technique, and thus achieves full diversity.

The simulation results of the proposed model demonstrate that the error performance of proposed system is better compared with that of the multiple-input multiple-output (MIMO) system on correlated as well as uncorrelated slow fading Rayleigh channels, when SNR is low. Given the transmission rate and the total transmission SNR, we demonstrate that there exists an optimal power allocation that minimizes the outage probability. We also show that on correlated slow fading Rayleigh channels, channel correlation will degrade the proposed system performance linearly in proportion to the correlation level.

The organization of this paper is as follows. Section \uppercase\expandafter{\romannumeral 2} provides literature review of related previous works. Section \uppercase\expandafter{\romannumeral 3} presents the cluster to VFC transmission model we used. Section \uppercase\expandafter{\romannumeral 4} outlines the outage probability analysis for the proposed system. Section \uppercase\expandafter{\romannumeral 5} provides the optimal power allocation strategy for the proposed system. Section \uppercase\expandafter{\romannumeral 6} discusses simulation results and analysis. Summary and conclusions are discussed in section \uppercase\expandafter{\romannumeral 7}.

\section{Literature Review}
The system model introduced in this paper includes two phases. The inter-cluster broadcast phase exploits cooperative diversity and the intra-cluster random beamforming phase exploits multiuser diversity. A remarkable amount of research has been carried out on both diversity techniques to improve the system performance.

Some earlier works \cite{Cover79} \cite{Sendonaris98} introduced a basic communication structure among nodes to exploit cooperative diversity. The results suggest that even in a noisy environment, the diversity created through cooperation between in-cluster nodes can not only increase the overall channel capacity, but also provide a more robust system to combat channel fading.

In most scattering environments, antenna diversity is practical, effective and, hence, a widely applied technique for reducing the effects of multipath fading. The classical approach is to use multiple antennas at the receiver and perform combining \cite{Brennan} or selection and switching \cite{Heath} in order to improve the quality of the received signal. Spatial diversity can also be achieved using space-time coding techniques \cite{Alamouti, Tarokh}. Compared with the low capacity and low reliability of single input and single output (SISO) system, multiple input and multiple output (MIMO) system provides higher capacity, better transmission quality, and larger coverage without increasing the total transmission energy. 

Laneman and Wornell developed cooperative diversity protocols considering the physical layer constraint. In this work, the spatial diversity achieved through coordinated transmission was exploit to combat multipath fading on a distributed antenna system \cite{Nicholas00} \cite{Wornell01} \cite{Wornell03}. In these works, they assume the nodes can transmit and receive simultaneously (full-duplex). They extended their work to include half-duplex transmission in \cite{Laneman04}. For a sensor network built in nonergodic scenarios like discrete-time channel models, it is more appropriate to use outage probability as a system performance metric \cite{Andrea01}. 

In addition to cooperative diversity, multiuser diversity can also increase channel capacity. Given the channel state information (CSI) of all users is known at VFC, through scheduling the best channel status to one user, the overall channel capacity can be increased \cite{Knopp95}. There are two constraint of this kind of multiuser diversity: all users are independent and there always exists a user having the best channel conditions. However, since VFC always tries to connect to the user with the best channel, the VFC may not be able to guarantee transmission quality to other users \cite{Laneman04}. In addition, it is not a fair channel assignment strategy.

In order to overcome the above problems, proportional fair scheduling algorithm \cite{Viswanath02} was developed to achieve a fair channel resource allocation. This technique, also known as random beamforming technique, assigns the channel resources based on the user feedback, so that the data rate as well as the overall throughput can be maximized. However, random beamforming technique only exploits diversity gain rather than multiplexing gain. Thus, in a limited bandwidth scenario, when a higher channel capacity is desired, random beamforming technique can not increase capacity linearly \cite{Chung03}. Therefore, some new techniques needed to developed to solve the problem.

Recently, some related works focus on combining cooperative diversity and multiuser diversity in relay networks \cite{Joung08} \cite{Zhang09} \cite{Chen09}. It has been proved that in relay networks, combine two diversity techniques can improve system performance \cite{Sun10}. In the proposed model, a combined diversity system model has been deployed in a cluster-based decentralized wireless sensor network. We improve the system performance in terms of outage probability through optimum power allocation.

\section{Communication Channel Model}
\begin{figure}
\begin{center}
\includegraphics[width=3.5in]{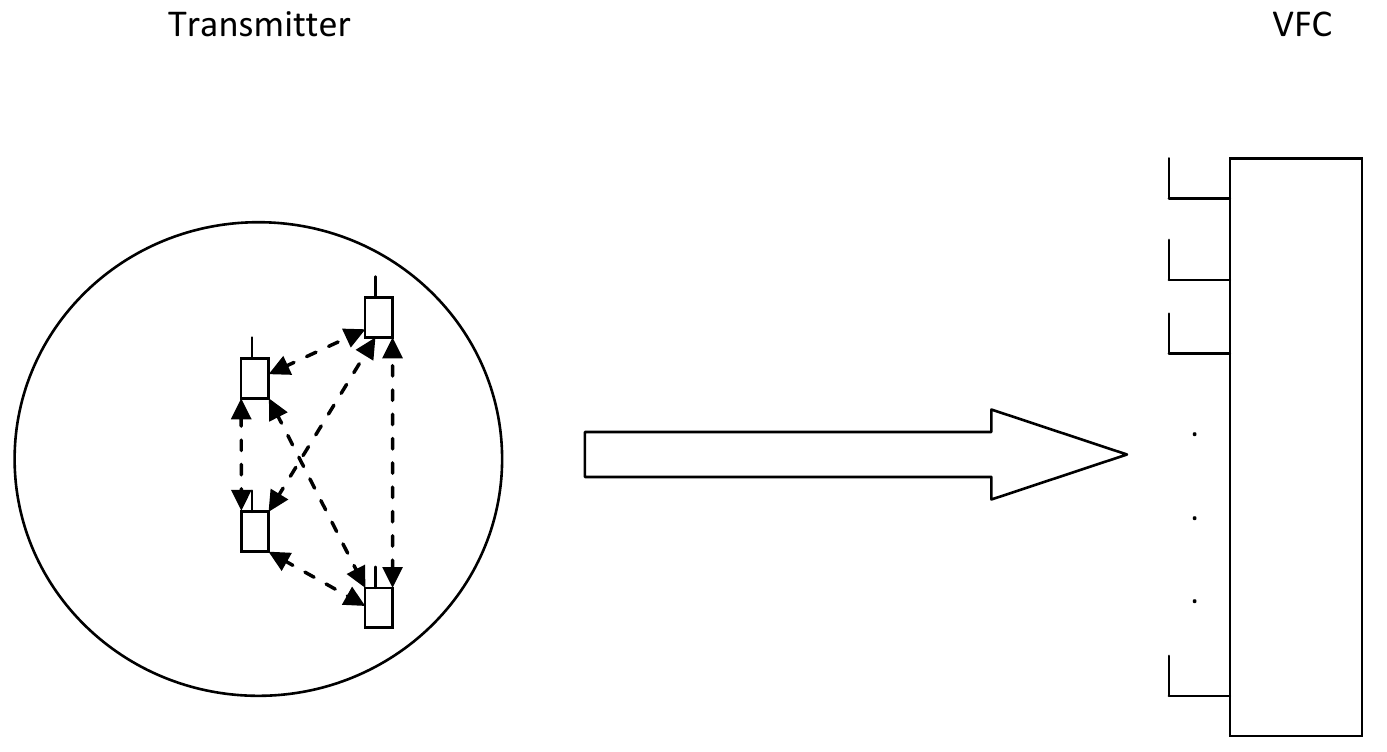}
\caption{Phase I: Intra-cluster broadcasting} \label{Fig:SYSMOD}
\end{center}
\end{figure}

\begin{figure}
\begin{center}
\includegraphics[width=3.5in]{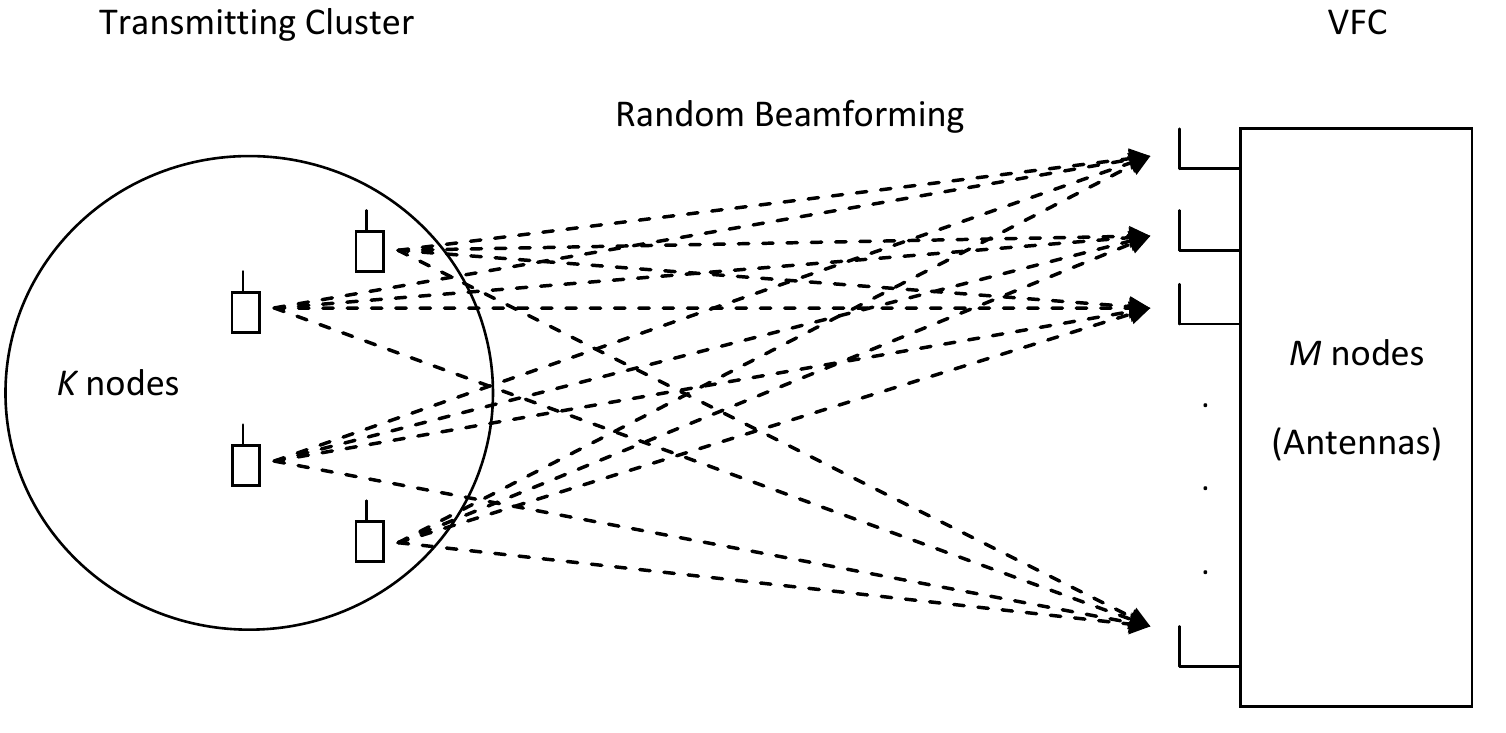}
\caption{Phase II: Random beamforming between a cluster and VFC} \label{Fig:TRANS}
\end{center}
\end{figure}

In this section, the channel models for the two phases of communication are discussed. Fig. \ref{Fig:SYSMOD} and Fig. \ref{Fig:TRANS} illustrate the broadcasting and random beamforming phases of transmission.

\subsection{Transmitter Side}
Assume that there are $\it{K}$
nodes each with one transceiver transmitting data to the VFC. Also, assume that the VFC is equipped with $\it{M}$ antennas.
Further, since sensor network is power-constrained, a reasonable assumption is that the total power allocated for all nodes for communicating their observations to the VFC is $P_{total}$. The transmission occurs in two phases as described below.

\subsubsection{Intra-cluster Broadcasting}
During the first phase, each sensor broadcasts its observations to all other nodes in the same cluster with certain power. Assume that all nodes decode the received data simultaneously \cite{Zhou06}. Half-duplexing transmission is assumed in this phase, where all the nodes in cluster can not send and receive at the same time on the same frequency. The total power used for all nodes to accomplish broadcasting (each node using one time unit to broadcast) is $P_{1}$. Therefore, the number of cooperative nodes, $\it{K}$, in each cluster depends on the selection of $P_{1}$.

\subsubsection{Random Beamforming between a Cluster and FC}
During this phase, all $K$ nodes in the cluster will transmit the data they aggregated to the VFC using total power $P_{2}$, and the complete transmission process takes one time unit. We propose a random beamforming technique during this phase with a transmit power for each cooperative node set to be $a_{i}$. We pre-multiply the input vector with a diagonal beamforming matrix $\bf{V_{b}}$ consisting of $K$ number of complex numbers $a_{i}exp(j\theta_{i})$, where $\theta_{i}\in[0,2\pi]$ for $i = 1,...,K$. The random beamforming matrix is designed such that ${a_{i}}~\scriptsize{\sim}~U(0,1)$, and $a_{i}$s are normalized such that $\sum_{i=0}^{K}{a_{i}^{2}} = P_{2}$. 

The total power ($P_{total}$) used for both phases is given by $P_{1} + P_{2}$. Based on this power constraint, the number of nodes $\it{K}$ in the cluster is selected such that it is proportional to $\frac{P_{1}}{P_{total}}$.

\subsection{Receiver Side}

The VFC receives data from each cluster over $K\times M$ random beamforming channel. In order to obtain full receiver diversity, we propose MRC technique to reconstruct the received data. For evaluating the performance of the random beamforming strategy, we use V-BLAST MIMO channel \cite{Wolniansky98} with the same amount of the total power as in random beamforming to compare, i.e.,
\begin{equation}
P_{MIMO} = P_{total}, \label{Eq:mimopower}
\end{equation}
where $P_{MIMO}$ represents the transmit power used for transmission over the MIMO channel. Since V-BLAST MIMO is a technique applied on MIMO channels as in Phase 2 of the proposed algorithm, we actually compare the performance of Phase 2 of the proposed algorithm with that of V-BLAST MIMO with the same power constraint and time duration.

\section{System Model and Error Probability}
\begin{figure}
\begin{center}
\includegraphics[width=3.5in]{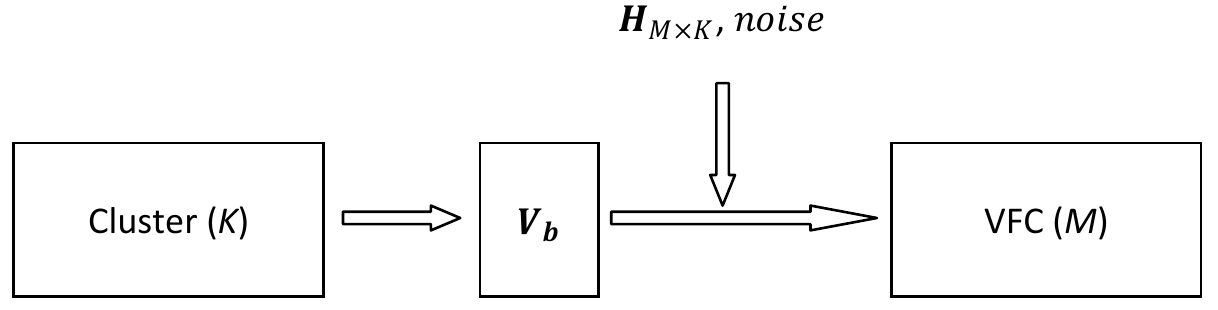}
\caption{Proposed system model: Communicating aggregated sensor observations over $M \times K$ random beamforming channel} \label{Fig:SYS}
\end{center}
\end{figure}

Before transmitting, the input is premultiplied by a random beamforming matrix $\bf{V_{b}}$. The system model we proposed is shown in Fig. \ref{Fig:SYS}.

The random beamforming communication model can be described by
\begin{equation}
\bf{y} = \bf{H} \bf{V_{b}} \bf{x} + \bf{n}, \label{Eq:rbchannel}
\end{equation}
where $\it{\bf{y}}$ stands for the channel output vector $[{{\it{y}}_{1}}, {{\it{y}}_{2}}, ... , {{\it{y}}_{M}}]^{T}$, $\it{\bf{x}}$ stands for the channel input vector $[{{\it{x}}_{1}}, {{\it{x}}_{2}}, ... , {{\it{x}}_{K}}]^{T}$, and $\it{\bf{n}}$ represents the channel noise vector, whose elements are assumed to be zero-mean white Gaussian with variance $\sigma_{n}^2$, i.e., ${\it{n}}_{i}~\scriptsize{\sim}~N(0, \sigma_{n}^2)$ $(i$ = $1,2, ... , m)$, $\it{\bf{H}}$ represents the channel fading coefficient matrix, and $\bf{V_{b}}$ represents random beamforming matrix. In comparison, a V-BLAST communication is assumed for MIMO model which is described by \cite{Wolniansky98}:
\begin{equation}
\bf{y} = \bf{H_{\tiny{MIMO}}} \bf{x} + \bf{n}. \label{Eq:virtual}
\end{equation}
\indent At VFC, the decoder implements MRC method to reconstruct the source, which is described by
\begin{equation}
\it{\bf{\hat{x}}} = {\bf{x}} + \frac{(\bf{H} \bf{V_{b}})^{H} \bf{n}}{||\bf{H} \bf{V_{b}}||^{2}}, \label{Eq:reconstruct}
\end{equation}
where $\it{\bf{\hat{x}}}$ represents the reconstructed source and ${(\bf{H} \bf{V_{b}})}^{H}$ represents the conjugate transpose of $\bf{H} \bf{V_{b}}$.

%, which is described by
%\begin{equation}
%\it{\bf{\hat{x}}} = {\bf{x}} + \frac{{(\bf{H} \bf{V_{b}})}^{H} \bf{n}}{||{\bf{H} \bf{V_{b}}}||}, \label{Eq:reconstruct}
%\end{equation}
%where $\it{\bf{\hat{x}}}$ represents the reconstructed source and ${(\bf{H} \bf{V_{b}})}^{H}$ represents the conjugate transpose of $\bf{H} \bf{V_{b}}$.

The signal-to-noise ratio (SNR) for the received signal can be described as \cite{Viswanath02}
\begin{equation}
SNR = \frac{P_{2}}{\sigma_{n}^{2}}||{\bf{H} \bf{V_{b}}}||^{2}, \label{Eq:snr}
\end{equation}
where $\sigma_{n}^{2}$ represents the variance of the channel noise.

Let us assume that the transmission rate between a cluster and VFC is $R_{tr}$. Then, according to Shannon's limit, when
\begin{equation}
R_{tr} \leq log_{2}(1+\frac{P_{2}}{\sigma_{n}^{2}}||{\bf{H} \bf{V_{b}}}||^{2}), \label{Eq:trs}
\end{equation}
the receiver is expected to decode the received data correctly. Unsatisfactory reception or outage occurs when this condition is not met.

We assume error-free transmission during broadcasting phase and focus on random beamforming phase. The system outage probability is given by
\begin{equation}
P_{out} = P(||{\bf{H} \bf{V_{b}}}||^{2} < \frac{2^{R_{tr}}-1}{\frac{P_{2}}{\sigma_{n}^{2}}}), \label{Eq:trlpr}
\end{equation}
which we use to as a performance measure for the proposed system. System optimization requires us to find the optimum power allocation strategy for the system to minimize this outage probability. The system performance depends on the number of transmitting nodes.

\section{Optimum Power Allocation Strategy}
In this section, we discuss an optimum power allocation strategy for the proposed system. The discussion is split into two parts: broadcasting and random beamforming.
\subsection{Broadcasting}
Assume that broadcast rate for all the nodes in the cluster are the same and can be represented as $R_{br}$. Each node will broadcast its data to all other nodes. According to Shannon limit, when
\begin{equation}
R_{br} \leq log_{2}(1+\frac{P_{1}}{K \sigma_{nbr}^{2}}), \label{Eq:brs}
\end{equation}
the receiver is expected to reconstruct the source correctly, where $\sigma_{nbr}^{2}$ represents noise on the broadcast channel. Eq (\ref{Eq:brs}) can also be written as,
\begin{equation}
P_{1} \geq K (2^{R_{br}}-1)\sigma_{nbr}^{2}, \label{Eq:brlan}
\end{equation}
which shows the lower bound on power used for broadcasting ($P_{1}$) for reliable transmission. We choose $P_{1}$ and $R_{br}$ such that they satisfying \ref{Eq:brlan}. Therefore, there is no outage during broadcasting.

\subsection{Random Beamforming}
Given two statistically independent random variables X and Y, the distribution of the random variable Z that is formed as the product $Z = XY$ \cite{Dale79}. Therefore, given that ${\bf{H}}~\scriptsize{\sim}~N(0,I)$ and ${\bf{V_{b}}}~\scriptsize{\sim}~U(0,I)$, ${\bf{H} \bf{V_{b}}} ~\scriptsize{\sim}~ PDF~of~H \times PDF~of~V_{b} = PDF~of~H$, thus, ${\bf{H} \bf{V_{b}}}~\scriptsize{\sim}~N(0,I)$, and $||{\bf{H} \bf{V_{b}}}||^{2}~\scriptsize{\sim}~\chi_{n_{t}}^{2}$ (Chi-Square distributed random varaible with $n_{t}$ degrees of freedom), where $n_{t} = M \times K$ \cite{jim}. Since the CDF of  $\chi_{n_{t}}^{2}$ is the regularized lower incomplete Gamma function $\gamma(.,.)$, the outage probability ($P_{out}$) can be described as \cite{Johnson95}
\begin{equation}
P_{out} = \frac{\gamma(\frac{MK}{2}, \frac{2^{R_{tr}}-1}{\frac{2 P_{2}}{\sigma_{n}^{2}}})}{\Gamma(\frac{MK}{2})}, \label{Eq:trlchi}
\end{equation}
where $\Gamma(.)$ is the Gamma function. A closed form expression for $P_{out}$ is too complex to derive. We used simulations for our optimization analysis.

\subsection{Optimization Problem}
The optimization problem is to minimize $P_{out}$ subject to the total power ($P_{total}$) constraint. Let us assume that  $P_{1} = \alpha P_{total}$ such that $P_{2} = (1-\alpha) P_{total}$. Then, the optimization problem can be expressed as follows:

\begin{equation}
\begin{array}{lc}
P^{*}_{out} =
\begin{array}{c}
min\\ \vspace{0.01cm}
\alpha,K
\end{array}
 P_{out},
\\s.t. ~ P_{1} + P_{2} = P_{total}, P_{1} = \alpha P_{total}, P_{2} = (1-\alpha) P_{total}
\end{array}  \label{Eq:tr2chi}
\end{equation}
where $P^{*}_{out}$ is the minimum outage probability. Assume the broadcast power for each node to maintain error-free broadcasting is $P_{s}$ and $M$ is a fixed number, then $P_{1} = K P_{s} = \alpha P_{total}$. Therefore, the relation between $\alpha$ and $\it{K}$ is given by  $ K = \alpha \frac{P_{total}}{P_{s}}$.

According to (\ref{Eq:trlchi}) and (\ref{Eq:tr2chi}), outage probability can be reduced by increasing the number of transmitting nodes ($K$), since increased number of transmitting nodes ($K$) will increase the multiuser diversity deployed in the random beamforming phase, which will reducing the outage probability. However, increasing $K$ will result in increased $\alpha$, which in turn, will reduce the power $P_{2}$ allocated for random beamforming. Reduced transmission power $P_{2}$ in the random beamforming phase will lead to decreased receiving SNR at FC, and that will degrade the transmission performance in outage probability. Therefore, increasing the power allocation factor ($\alpha$) will introduce more multiuser diversity and deduct transmission power in random beamforming at the same time, and there exists an optimal power allocation that minimizes the outage probability. In the following section, numerical results are provided to illustrate the effect of $\alpha$.

%Then, $P^{*}_{out}$ can be represented as,
%\begin{equation}
%P^{*}_{out} = \frac{\gamma(\frac{\it{B_{\tau}}}{2}, \frac{2^{R_{tr}}-1}{\frac{2(1-\alpha_{\tau})P_{total}}{\sigma_{n}^{2}}})}{\Gamma(\frac{\it{B_{\tau}}}{2})}. \label{Eq:final}
%\end{equation}
%Therefore, the optimal power allocation is,
%\begin{equation}
%P_{1} = \alpha_{\tau} P_{total}, P_{2} = (1-\alpha_{\tau})P_{total} \label{Eq:optimal}
%\end{equation}

\section{Performance Evaluation}
The proposed model is simulated in MATLAB and its performance is evaluated by estimating the outage probability on slow fading correlated and uncorrelated Rayleigh channels. All nodes are assumed to be uniformly distributed in each cluster, so that the broadcast power $P_{1}$ is directly proportional to the number of cooperating nodes. Also, $\alpha$ is  varied within the range (0.2 to 0.8) such that the transmission quality is maintained, and $\frac{P_{total}}{P_{s}}$ is assumed to be 15, so that $ K = 15 \alpha$. For all simulations, broadcasting rate $R_{br}$ is set to 2 bits/Hz and random beamforming $R_{tr}$ = 3 bits/Hz.

\subsection{Slow Fading Uncorrelated Rayleigh Channel}
Consider the case of slow fading uncorrelated Rayleigh channel in which the channel gains remain constant for each use of the channel. The channel fading states are modeled as independent and identically distributed zero mean and unit variance complex Gaussian random variables. In order to implement random beamforming, the base station only needs to know the overall SNR, which is defined as $\frac{P_{total}}{\sigma_{n}^{2}}$. To achieve multiuser diversity, we need to add fast time-scale fluctuations on the channel using ${\bf{V_{b}}}$.

\begin{figure}
\begin{center}
\includegraphics[width=3.5in]{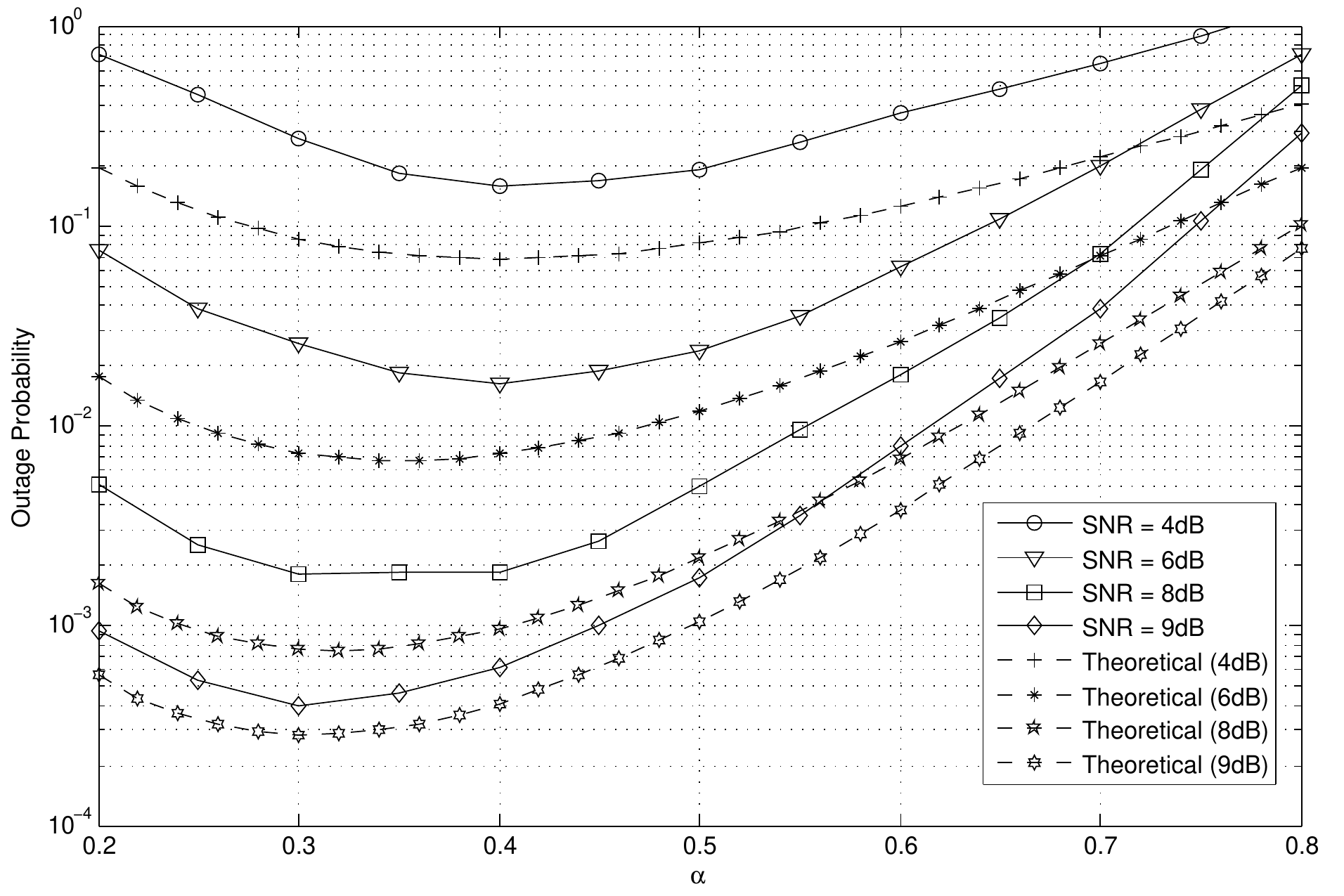}
\caption{Outage probability as a function of the power allocation factors for different channel SNRs} \label{Fig:extreme}
\end{center}
\end{figure}

As shown in the Fig. \ref{Fig:extreme}, when SNR is equal to 4 dB, $P_{out}$ reaches the minimum at $\alpha_{\tau} = 0.4$. As SNR is increased, the value of $\alpha_{\tau}$ gets reduced. For instance, when SNR is equal to 9 dB, $P_{out}$ achieves the minimum  at $\alpha_{\tau} = 0.3$. Also,please note that the theoretical value is a lower bound of the outage.

\begin{figure}
\begin{center}
\includegraphics[width=3.5in]{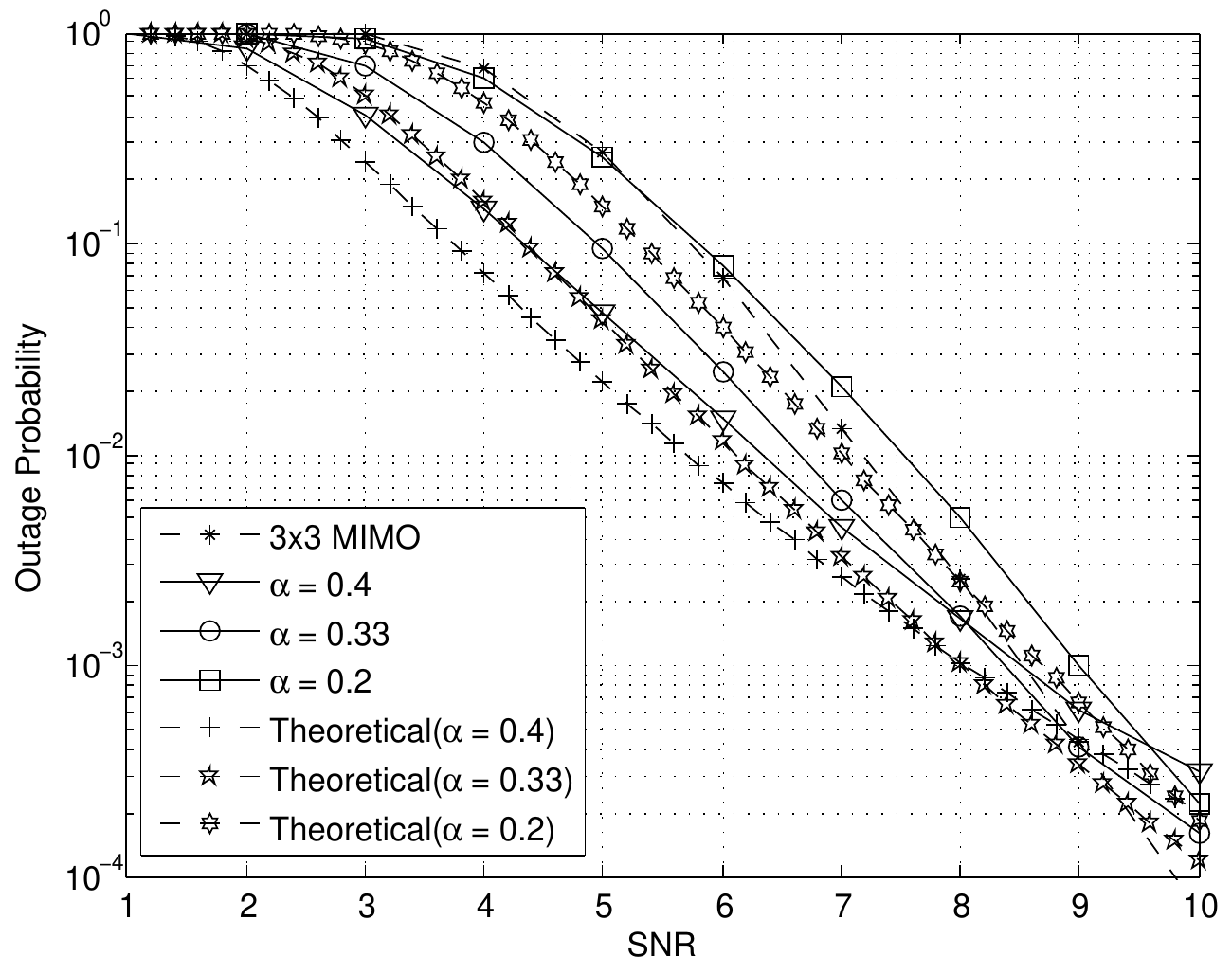}
\caption{Outage probability as a function of the channel SNR for different power allocations} \label{Fig:slow}
\end{center}
\end{figure}

As shown in the Fig. \ref{Fig:slow}, when SNR is equal or lower than 9 dB, the proposed system performs better compared with the 3x3 MIMO \cite{Wolniansky98} in terms of outage probability, where $\it{K}$ equals to 3, 5, and 6.

Fig. \ref{Fig:slow} also demonstrates that when SNR is lower than 8dB, the proposed system exhibits the lowest outage probability when $\alpha_{\tau}$ = 0.4. When SNR is higher than 8dB, $\alpha_{\tau}$ gets reduced to 0.3.

\subsection{Correlated Rayleigh Fading Channel}
\begin{figure}
\begin{center}
\includegraphics[width=3.5in]{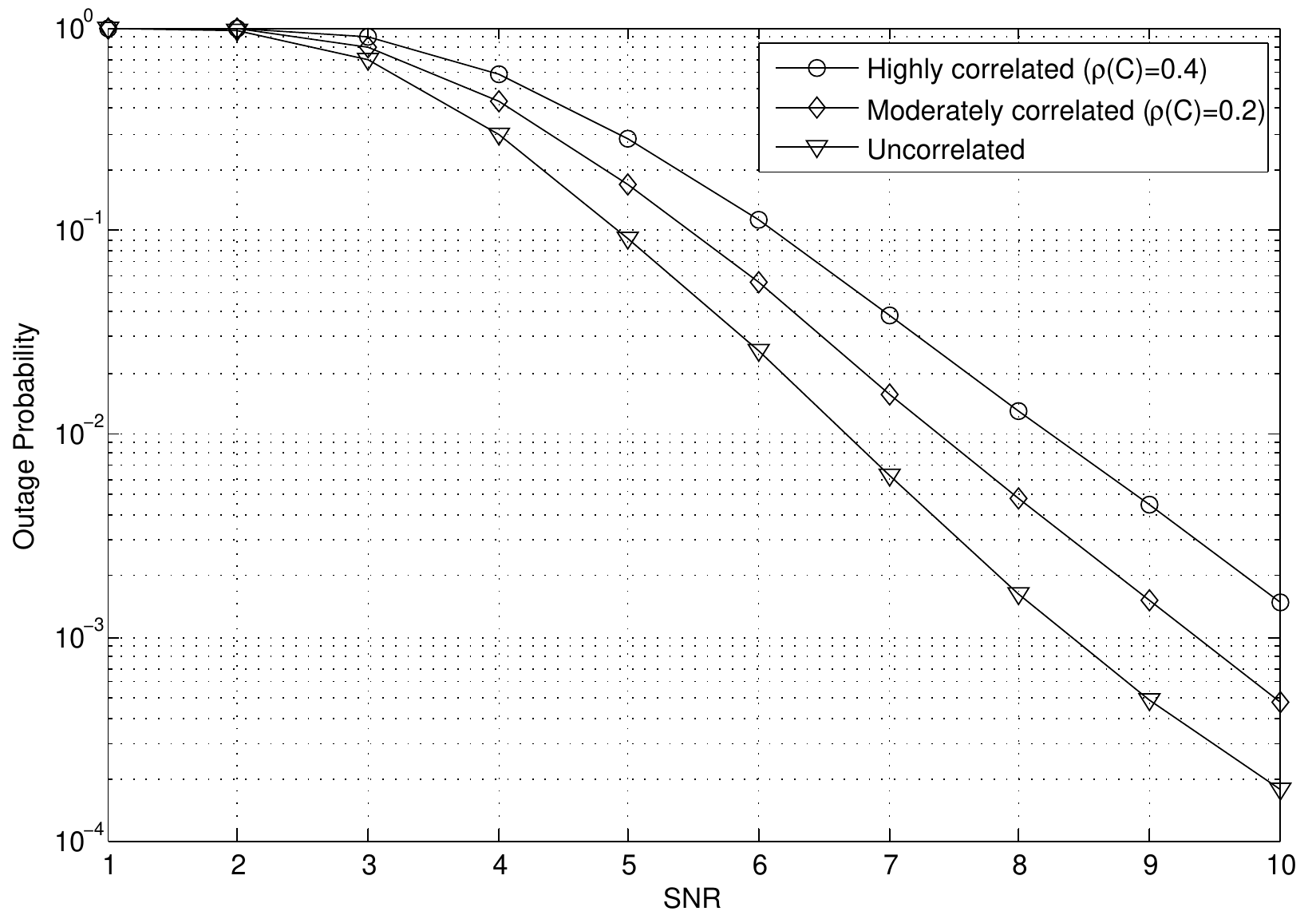}
\caption{Outage probability as a function of the received SNR on correlated and uncorrelated Rayleigh fading channels with 9 transmitting antennas and $\alpha = 30\%$} \label{Fig:fast}
\end{center}
\end{figure}

To simulate the correlated Rayleigh fading environment, we multiplied the correlation matrix \textbf{C} with the channel matrix $\bf{H}$. The correlation level ($\rho({\textbf{C}})$) is given by
\begin{equation}
\rho({\textbf{C}}) = \frac{||{\textbf{C}}-diag({\textbf{C}})||_{F}}{||diag({\textbf{C}})||_{F}}, \label{Eq:corrlevel}
\end{equation}
where $||{\textbf{C}}||_{F}$ is the Frobenius norm of \textbf{C} matrix and $diag({\textbf{C}})$ is the matrix that contains only the diagonal components of \textbf{C}.
\indent The correlated Rayleigh fading channel can be described by
\begin{equation}
\bf{y} = \bf{C} \bf{H} \bf{V_{b}} \bf{x} + \bf{n}. \label{Eq:corrchannel}
\end{equation}
Assuming that the eigenvalues of the channel matrix $\bf{V_{b}}^{H} \bf{H}^{H} \bf{C}^{H} \bf{C} \bf{H} \bf{V_{b}}$ are $\lambda_{1}, \lambda_{2}, ... \lambda_{n}$, the channel capacity is linearly proportional to their product $\lambda_{1}\lambda_{2}...\lambda_{n}$. Channel correlation increases the condition number of the effective channel matrix $\bf{V_{b}}^{H} \bf{H}^{H} \bf{C}^{H} \bf{C} \bf{H} \bf{V_{b}}$ \cite{Shiu00} by increasing the difference between the largest and smallest eigenvalues. Hence, the channel capacity decreases with increased $\rho({\textbf{C}})$.   This is demonstrated in Fig. \ref{Fig:fast}.

\section{Conclusions}

In this paper, we proposed a cluster based cooperative communication strategy using random beamforming technique. In the first place, through comparing the outage performance between the proposed system and V-BLAST MIMO, we demonstrated that when the received SNR is low, the outage probability of the proposed system is better compared with that of the multiple-input and multiple-output system on slow fading Rayleigh fading channels.

Through analyzing the optimal power allocation factor ($\alpha$) under different overall SNRs, we find out that: (1) when the overall SNR is low, e.g. lower than 8dB, increased multiuser diversity ($\alpha = 0.4$, $P_{1} = 0.4 P_{total}$, $P_{2} = 0.6 P_{total}$) show greater advantage in combating the channel fading and noise compared with increasing transmission power in random beamforming phase ($\alpha = 0.3$, $P_{1} = 0.3 P_{total}$, $P_{2} = 0.7 P_{total}$); (2) when the overall SNR is high, e.g. higher than 8dB, increased transmission power in random beamforming phase ($\alpha = 0.3$) will show more benefits in reducing the outage probability than increased multiuser diversity ($\alpha = 0.4$).

We also compare the performance on correlated and uncorrelated Rayleigh fading channels. Our results demonstrate that since the channel capacity decreases with increased correlation level $\rho({\textbf{C}})$, channel correlation will degrade the proposed system outage performance in linearly proportional to the correlation level.

%\singlespacing
\bibliographystyle{IEEEtran}
\bibliography{new}

\end{document}